# Dynamics of Polar Skyrmion Bubbles under Electric Fields

Ruixue Zhu[1, ‡], Zhexin Jiang[2, ‡], Xinxin Zhang[1, ‡], Xiangli Zhong[3, ‡], Congbing Tan[4, *], Mingwei Liu[4], Yuanwei Sun[1], Xiaomei Li[1,5], Ruishi Qi[1], Ke Qu[1], Zhetong Liu[1], Mei Wu[1], Mingqiang Li[1], Boyuan Huang[7], Zhi Xu[5], Jinbin Wang[3], Kaihui Liu[9, 10], Peng Gao[1, 10, *], Jie Wang[2, 8, *], Jiangyu Li[6] and Xuedong Bai[5, *]

[1]Electron Microscopy Laboratory and International Center for Quantum Materials, School of Physics, Peking University, Beijing 100871, China

[2]Department of Engineering Mechanics, Zhejiang University, Hangzhou 310027, Zhejiang, China

[3]School of Materials Science and Engineering, Xiangtan University, Xiangtan 411105, Hunan, China

[4]Hunan Provincial Key Laboratory of Intelligent Sensors and Advanced Sensor Materials, School of Physics and Electronics, Hunan University of Science and Technology, Xiangtan 411201, Hunan, China

[5]Beijing National Laboratory for Condensed Matter Physics, Institute of Physics, Chinese Academy of Sciences, Beijing 100190, China

[6]Guangdong Provincial Key Laboratory of Functional Oxide Materials and Devices, Southern University of Science and Technology, Shenzhen 518055, Guangdong, China

[7]Department of Materials Science and Engineering, Southern University of Science and Technology, Shenzhen 518055, Guangdong, China

[8]Key Laboratory of Soft Machines and Smart Devices of Zhejiang Province, Zhejiang University, Hangzhou 310027, Zhejiang, China

[9]State Key Laboratory for Artificial Microstructure & Mesoscopic Physics, School of Physics, Peking University, Beijing 100871, China

[10]Collaborative Innovation Centre of Quantum Matter, Beijing 100871, China

[‡]These authors contributed equally to the work.

* Corresponding authors. Email: cbtan@xtu.edu.cn; p-gao@pku.edu.cn; jw@zju.edu.cn; xdbai@iphy.ac.cn

**Abstract:** Room-temperature polar skyrmion bubbles that are recently found in oxide superlattice, have received enormous interests for their potential applications in nanoelectronics due to the nanometer size, emergent chirality, and negative capacitance. For practical applications, the ability to controllably manipulate them by using external stimuli is prerequisite. Here, we study the dynamics of individual polar skyrmion bubbles at the nanoscale by using in situ biasing in a scanning transmission electron microscope. The reversible electric field-driven phase transition between topological and trivial polar states are demonstrated. We create, erase and monitor the shrinkage and expansion of individual polar skyrmions. We find that their transition behaviors are substantially different from that of magnetic analogue. The underlying mechanism is discussed by combing with the phase-field simulations. The controllable manipulation of nanoscale polar skyrmions allows us to tune the dielectric permittivity at atomic scale and detailed knowledge of their phase transition behaviors provides fundamentals for their applications in nanoelectronics.

**Introduction**

Topological polar structures, such as vortices(1-6), antivortices(7, 8), flux-closures(9-12), skyrmion bubbles(13, 14), merons(15), toroidal topological textures(16) that are the electric analogues of magnetic topologies, have been discovered in low-dimensional ferroelectrics and oxide superlattices in the past few years(17-19). These particle-like topological defects are stabilized by the complex interplay of elastic, electrostatic and gradient energies, and exhibit exotic properties, such as negative capacitance(20, 21) and chirality(13, 22-24), which make them promising candidates for robust, ultralow-power-consumption and high-density nanoelectronic devices in the post-Moore era.

Most of these applications require the ability to controllably manipulate the polar states and thus the knowledge of their phase transition behaviors under the external stimuli. Since in polar systems the dielectric anisotropy is usually much stronger than the magnetic one, the formation energy is usually large and the phase transition behavior of polar topologies are substantially different from those in the magnetics. Several theoretical and experimental reports have studied the phase transition pathway of polar topological defects under various external stimuli. For example, an electric field or stress can break the balance between depolarization field and strain boundary condition, thus flux-closures and vortices can be reversibly converted between topological and trivial ferroelectric states(25-29). In addition, a terahertz-field can stimulate the collective dynamics of polar vortices(30).

In the case of polar skyrmions, heating can certainly achieve a topological phase transition according to Landau theory(31), and even form an intermediate state, the meron, with the topological number halved(32). Furthermore, theoretical works(33-35) have predicted that the phase transition between topological polar skyrmions and trivial ferroelectric phase can take place even at room temperature, which is highly desired for most of applications in nanoelectronics. Experimentally, the macroscopic dielectric measurements and X-ray diffraction reciprocal space mappings have detected the response of polar skyrmions to electric field and observed the phase transition between topological and trivial states(20), during which a large tunability of the dielectric permittivity was achieved. However, for practical applications, it is necessary to know how the individual polar skyrmions evolves and how they interact with each other under external fields, which motivated the present study of probing the phase transition behavior at nanoscale in real-space.

In this work, we directly observe the electric field-driven evolution of individual polar skyrmions by using in situ scanning transmission electron microscopy (STEM) technique. We find that polar skyrmion bubbles can be created via fragmentation of the pristine stripe domains under an electric field. Further increasing the field leads to their shrinkage and annihilation to form trivial monodomain, during which the smaller ones disappear first, while an opposite bias causes the expansion and merger into monodomain. Upon returning to zero bias, isolated skyrmion bubbles re-nucleate within the monodomain. Notably, in the magnetic systems the electrical current mainly drives the motion of magnetic skyrmions and their shape remains almost unchanged. For the polar skyrmions under the electric field, the expansion, shrinkage, disappearance and nucleation dominate the phase transition, while no significant lateral motion is observed. The demonstration of controllable manipulation of skyrmion-stripe-monodomain switching allows us to tune dielelctric permittvity on the atomic scale via electric fields and the revealed phase transition behaviors provide necessary information for the design of nanodevices based on polar skyrmion bubbles.

**Results and Discussion**

The low-magnification medium angle annular dark field (MAADF)-STEM image in Fig. 1a shows a $(SrTiO_3)_{11}/(PbTiO_3)_{22}/(SrTiO_3)_{11}$ trilayer film grown on $SrTiO_3$ (STO) (001) single crystal substrate by pulsed laser deposition (PLD), with an $SrRuO_3$ (SRO) conductive layer in the middle serving as the bottom electrode (Methods). Both the strain contrast in cross-sectional MAADF image in Fig. 1a and diffraction contrast in the dark field image in Fig. 1b (and Supplementary Fig. 1) show non-uniform contrast in the PTO layer from the cross-sectional view, which originate from anti-parallel out-of-plane polarization states as illustrated in the simulated polarization mapping (Fig. 1c). The planar-view image (Fig. 1d) further suggests the existence of polar skyrmion bubbles coexisting with elongated stripe domains, well consistent with previous studies(13). Such a domain structure can also be well reproduced by the phase-field simulations shown in Fig. 1e and Fig. 1f, where the color indicates the direction of in-plane polarization and the saturation of color is proportional to the polarization magnitude. The configuration of three-dimensional electric dipoles distribution for a single polar skyrmion is displayed in Supplementary Fig. 2.

To reveal the dynamic evolution of skyrmion bubbles and elongated stripes under an applied electric field, we use a tungsten probe and the conductive SRO layer as electrodes to apply an out-of-plane electric field to the film in the electron microscope, as schematically shown in Fig. 2a. In situ experiments were performed in STEM mode slightly off [001] axis for better contrast. Several typical snapshots of the dynamic evolution selected from Supplementary Movie 1 are presented in Fig. 2b. We first apply a downward positive bias to the trilayer, which is anti-parallel to the polarization of the domain cores. Upon increasing the bias from 0 V to +6 V, the elongated stripe domains shrink and shorten thereby becoming thinner and smaller (a few typical domains outlined in purple and red). Some of stripe domains are fragmentized into isolated skyrmion bubbles. A subsequent transition from these newly formed skyrmion bubbles and residual stripe domains to the monodomain occurs as the positive bias increases further. In our study, the critical bias for the topological phase transition is estimated to be ~10 V, although a quantitative estimation of actual electric field distribution is difficult due to the complicated contact geometry. However, such an electric field-stabilized uniform downward-poled ferroelectric monodomain is metastable. After the positive bias is removed, the isolated skyrmion bubbles returns back spontaneously.

Continuing to apply an upward negative bias to the skyrmion bubble state, this driving force widens skyrmion bubbles and elongated stripes, and eventually coalesced into an upward-poled monodomain. Again, removal of the bias leads to regeneration of mixed domains similar to the initial state. The evolutions of domain structures during the full circles demonstrate the reversibility and repeatability of phase transitions between topological and trivial polar states.

The dynamic evolution of the coexisting topological domains is also simulated by applying a homogeneous vertical electric field in phase-field simulations to reveal the possible intermediate polarization states during the transition process and explore the switching mechanisms. The in-plane polarization evolution in Fig. 3 verifies that stripe domains gradually shrink and some of them break into skyrmion bubbles as the electric field increases (white dashed box), in nice agreement with the experimental results above. Topological phase transition occurs under a critical electric field of $E_{[001]} = 3400$ kV/cm, whereupon all the topological domains vanished and saturated into a uniform monodomain state, where the electric dipoles in PTO layer are aligned along the same direction as the applied electric field. Similar simulated evolution under an opposite electric field is shown in Supplementary Fig. 3, in which the stripe domain exhibiting divergent in-plane polarization gradually widens and undergoes a topological phase transition at the critical electric field of $E_{[001]}$ = -3400 kV/cm.

We also probe the phase transition behavior of polar skyrmions in $[(SrTiO_3)_{11}/(PbTiO_3)_{22}/(SrTiO_3)_{11}]_6$ superlattices, which is expected to possess higher density of skyrmion bubbles according to the previous study(13). Indeed, the MADDF image of the cross-section in Fig. 4a, the corresponding dark field image in Supplementary Fig. 4, and the planar-view STEM image in Fig. 4b confirm the existence of higher density of skyrmions and that stripe domains are shorter and less. The topological phase transition from skyrmion bubbles to monodomain is observed as the bias increases from 0 to 6 V (Fig. 4c), in agreement with the previous report(20). Some smaller isolated skyrmions start to disappear even at a bias of 3 V (indicated by cyan arrows), while large ones gradually shrink first (indicated by purple arrows) before vanishing. Removal of the external stimuli leads to spontaneous reappearance of skyrmion bubbles. Similarly, further decreasing the bias to a negative value, we find that skyrmion bubbles expand and merge into elongated stripe domains by switching the polarization component opposite to the electric field. Again, no significant lateral motion is observed during the phase transition, although the regenerated position and shape of skyrmion bubbles can be slightly different from the original ones (see Supplementary Movie 2 for details).

Thus far, our experiments mainly reveal two phenomena. The first is the electric field-driven generation and erasure of polar skyrmion bubbles. Secondly, the morphology and size of topological textures can be controllably manipulated by external bias, which is manifested in the conversion between skyrmion bubbles and elongated stripes. The evolution of these skyrmions is independent, rather than collective polarization dynamics behavior in polar vortices(30). The above experimental observations in real space can be mutually verified with the recently published simulation results(20, 34, 35). To further understand the evolution process and the underlying mechanism, we compare the approach of generating polar skyrmions through fragmenting stripe domains with the case in magnetic system(36). Different from the spin-orbit torque in magnets, the external electric field which destroys the electrostatic balance in ferroelectrics is the driving force of the expansion or shrinkage of topological structures via the motion of 180° domain walls. Reminiscent of the Rayleigh instability of fluids, the original shape of stripe domains tends to divide into circular skyrmion bubbles to minimize surface area under external electric field, which considered as "nanodomains ejection from domain walls" in previous report(37). Further increase the electric field leads to the topological phase transition, at which the electric field as the driving force crosses the finite energy barrier between different topological states. It is worth mentioning that the motion of magnetic skyrmions can be driven by the current induced spin-torque(38, 39). However, no obvious moving events have been observed in ferroelectric system, being substantially different from the magnetic analogue. We speculate that even if the polar skyrmions move, their amplitudes are very small, and the polar skyrmions have a high probability of being erased before this movement is observed.

The dynamic manipulation of polar skyrmions is accompanied by the change of dielectric properties. Das *et al.*(20) revealed that the skyrmions walls have negative permittivity and are more susceptible to external electric field, thus provide a guideline for tuning dielectric permittivity and our investigations complement the microscopic lack. The macro dielectric permittivity measurements of PTO/STO trilayer and superlattice both show an asymmetric curve in response to opposite electric fields. It can be well understood from the perspective of our in situ experiments at the nanoscale, i.e., the application of a positive electric field is accompanied by the shrinkage of polar skyrmions, and the volume of the domain wall with negative permittivity decreases, resulting in a rapid drop of total dielectric permittivity. Conversely, a negative electric field causes the domain wall to expand, thereby slowing the decrease in total permittivity. Therefore, enriched microstructure information lays the foundation for the interpretation of macroscopic performance and provides guideline for design related applications.

In summary, our work has experimentally revealed electrically driven phase transition behaviors of polar skyrmion bubbles in $SrTiO_3$/$PbTiO_3$/$SrTiO_3$ trilayer and superlattice at nanoscale. By applying an electric field with proper direction and magnitude, we can create, erase, and change the size and shape of polar skyrmion bubbles, as well as realizing the conversion between skyrmion bubbles, stripe domains and trivial ferroelectric monodomains. The phase transition behaviors are very different from the magnetic counterpart. This deterministic electric field control of topological solitons opens the door for their applications in polar skyrmion-based nanoelectronics.

**Acknowledgements**

This research was supported by the National Natural Science Foundation of China (11974023, 11972320, 52021006), the Key R&D Program of Guangdong Province (2018B030327001, 2018B010109009, 2020B010189001, 2019B010931001), the National Equipment Program of China (ZDYZ2015-1), and the "2011 Program" from the Peking-Tsinghua-IOP Collaborative Innovation Center of Quantum Matter, the Pearl River Talent Recruitment Program of Guangdong Province (2019ZT08C321). And we acknowledge the support of Guangdong Provincial Key Laboratory Program (2021B1212040001) from the Department of Science and Technology of Guangdong Province. We also thank Electron Microscopy Laboratory in Peking University for the use of the Cs-corrected electron microscope.

**Methods**

**Fabrication of trilayer and superlattice.** $(SrTiO_3)_{11}/(PbTiO_3)_{22}/(SrTiO_3)_{11}$ trilayer and $[(SrTiO_3)_{11}/(PbTiO_3)_{22}/(SrTiO_3)_{11}]_6$ superlattice were deposited on 12 nm SRO-buffered (001)-oriented STO single-crystal substrate via pulsed-laser deposition (PVD-5000). The sintered PTO (10 mol% Pb-enriched) and STO targets were used for PTO and STO film deposition. The substrate temperature and oxygen pressure for the depositions of the PTO and STO film were 600 °C and 200 mTorr, respectively. The laser energy for the film deposition was set to 360 mJ pulse$^{-1}$, which was crucial for the presence of polar skyrmions in the fully strained PTO layer. Thicknesses of the PTO and STO layers were controlled to be 22- and 11-unit cells using deposition time. After deposition, the films were slowly cooled down to room temperature at an oxygen pressure of 200 mTorr.

**TEM sample preparation.** The planar view samples for in situ experiments were prepared by mechanical polishing (thickness is less than 20 µm) and subsequent argon ion milling with PIPS$^{TM}$ (Model 691, Gatan Inc.). We used low voltage (0.5 kV) and high angle during milling to reduce damage and surface amorphous layer. Cross-sectional samples were prepared by focused ion beam (Helios G4) etching.

**Electron microscopy characterization.** Planar view STEM images were recorded at 300 kV using an aberration-corrected FEI Titan Themis G2. The beam convergence angle for low-magnification STEM imaging is 17 mrad and the collection angle is 20-123 mrad for better contrast. MAADF-STEM images of the cross-section samples were required with collection angle in the range of 31-191 mrad. Corresponding dark field transmission electron microscopy images were collected under the two-beam condition with **g** = (200), operated at 200 kV in an FEI F20 microscope. In situ experiments were carried out on an aberration-corrected FEI Titan Themis G2 at 300 kV in STEM mode. The electrical holder we used is customed from ZepTools Technology Company, with high stability. An electrochemically etched tungsten tip serves as the positive or negative electrode of electrical switching, and the SRO conductive layer connected to the holder ground serves as the other electrode. The STEM images were recorded with 4,096 × 4,096 pixels for each frame and a dwell time of 1 µs per pixel.

**Phase-field simulation.** In order to study the polar skyrmion bubbles and stripes in the ferroelectric trilayer, a phase-field model of STO/PTO/STO is employed to simulate the topological structures and their evolution under external electric field. In the phase-field model, the total free energy $F$ is the sum of Landau free energy, domain wall

energy, elastic energy and electric energy. The total free energy has the following form(40):

$$F = \int [\alpha_i P_i^2 + \alpha_{ij} P_i^2 P_j^2 + \alpha_{ijk} P_i^2 P_j^2 P_k^2 + \frac{1}{2} g_{ijkl} \left( \frac{\partial P_i}{\partial x_j} \frac{\partial P_k}{\partial x_l} \right) + \frac{1}{2} c_{ijkl} \varepsilon_{ij} \varepsilon_{kl}$$
$$- q_{ijkl} \varepsilon_{ij} P_k P_l - \frac{1}{2} \varepsilon_0 \varepsilon_r E_i E_i - E_i P_i] dV \tag{1}$$

in which $\alpha_i$, $\alpha_{ij}$ and $\alpha_{ijk}$ denote the Landau coefficients, $P_i$, $\varepsilon_{ij}$ and $E_i$ represent the polarization, strain and electric field, $g_{ijkl}$ is the gradient energy coefficient, $c_{ijkl}$ is the elastic constant, $q_{ijkl}$ is the electrostrictive coefficient and $\varepsilon_0$ is the vacuum permittivity. As for $\varepsilon_r$, the relative dielectric constant of the background material, is set to 66 and 300 for PTO and STO respectively. All of the values for the required coefficients at room temperature mentioned above can be found in the previous articles(41-43). The repeating subscripts in equation (1) imply summation over the Cartesian coordinate components $x_i$ ($i = 1, 2$ and $3$).

In this study, the temporal evolution of the domain structure is obtained by solving the time-dependent Ginzburg-Landau (TDGL) equations (44):

$$\frac{\partial P_i(\mathbf{r}, t)}{\partial t} = -L \frac{\delta F}{\delta P_i(\mathbf{r}, t)} \tag{2}$$

where $L$, $\mathbf{r}$, and $t$ denote the kinetic coefficient, spatial position vector, and time. The TDGL equations choose the polarization vector $\mathbf{P} = (P_1, P_2, P_3)$ as the order parameter.

In addition, both the mechanical equilibrium equation:

$$\sigma_{ij,j} = \frac{\partial}{\partial x_j} \left( \frac{\partial f}{\partial \varepsilon_{ij}} \right) = 0 \tag{3}$$

and the Maxwell's equation:

$$D_{i,i} = -\frac{\partial}{\partial x_i} \left( \frac{\partial f}{\partial E_i} \right) = 0 \tag{4}$$

are satisfied at the same time, where $\sigma_{ij}$ and the $D_i$ are the stress and electric displacement components. In this work, the semi-implicit Fourier-spectral method is employed to numerically solve the partial differential equations(45).

The size of the simulation box is $280 \times 280 \times 44$, corresponding to the real-space size of $140 \times 140 \times 17.6$ nm³, where the thickness of the PTO and STO is 8.8 and 4.4 nm. Periodic boundary conditions are assumed in the two in-plane dimensions, whereas a superposition scheme is applied in the thickness dimension. To obtain the distribution of inhomogeneous depolarization field, short-circuit electric boundary conditions are applied to the top and bottom of the trilayer. On this basis, an external uniform electric field is used to induce the whole evolution of polar skyrmion bubbles and stripes. Small random fluctuation (<$0.01P_0$, where $P_0 = 0.757$ $Cm^{-2}$ is the spontaneous polarization of PTO at room temperature) is used as the initial values of polarization to initiate the polarization evolution. The gradient coefficients are assumed to be isotropic with $g_{11} = 1.730 \times 10^{-11}$ and $g_{44} = 0.865 \times 10^{-11}$ $m^4NC^{-2}$.

**Figures and captions**

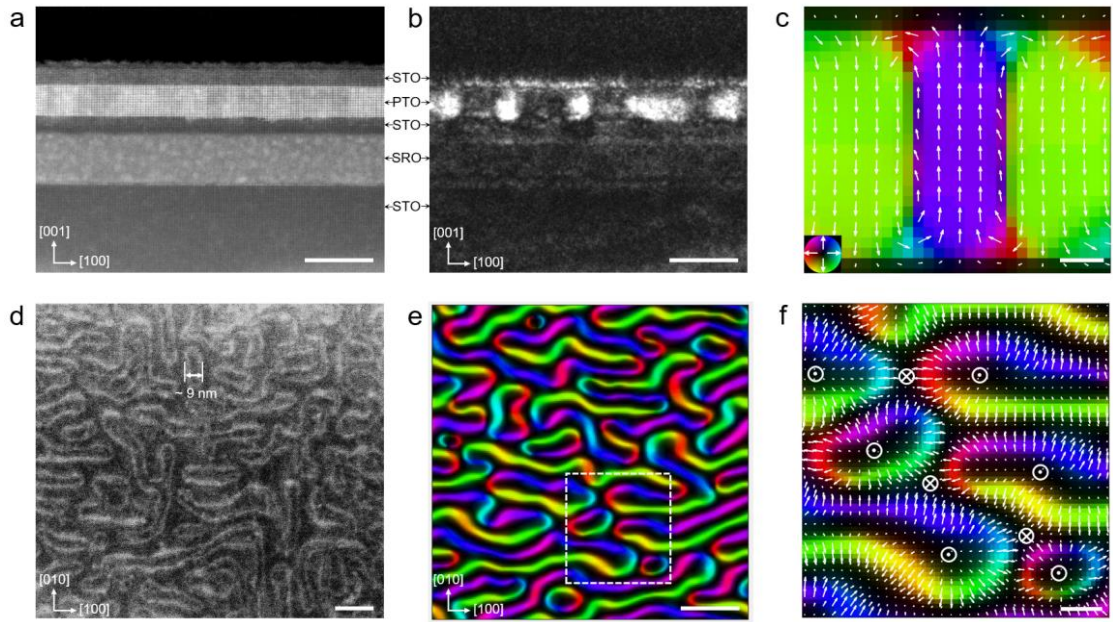

**Fig. 1. Observation of mixed polar structures of skyrmion bubbles and stripe domains. a**, **b**, Cross-sectional STEM-MAADF image of a trilayer sample $(SrTiO_3)_{11}/(PbTiO_3)_{22}/(SrTiO_3)_{11}$ (a), and a dark field TEM image (b), showing a contrast of alternating bright and dark regions within PTO layer. Scale bars, 20 nm. **c**, Antiparallel polarization distribution of cross sectional view from phase-field simulations. The white arrows denote the direction and magnitude of polarization, and the background color mapping represents the polarization direction.The saturation of the color is proportional to the polariztaion magnitude. Scale bar, 2 nm. **d**, Low-magnification STEM image of planar view of trilayer showing a coexisting structure of skyrmion bubbles and stripe domains. Scale bar, 20 nm. For these domains, dark contrast regions possessing out-of-plane polarization (upward or downward) are separated by bright contrast domain walls with in-plane polarization. **e, f**, Low-magnification (e) and enlarged view of the white box (f) color mappings constructed from phase-field simulations, depicting in-plane polarization distribution with hedgehog-like feature. The out-of-plane polarization in the dark regions is represented by the symbols '⊗' (pointing into the page) and '⊙' (pointing out of the page). The scale bar in (e) is 20 nm and that in (f) is 2 nm.

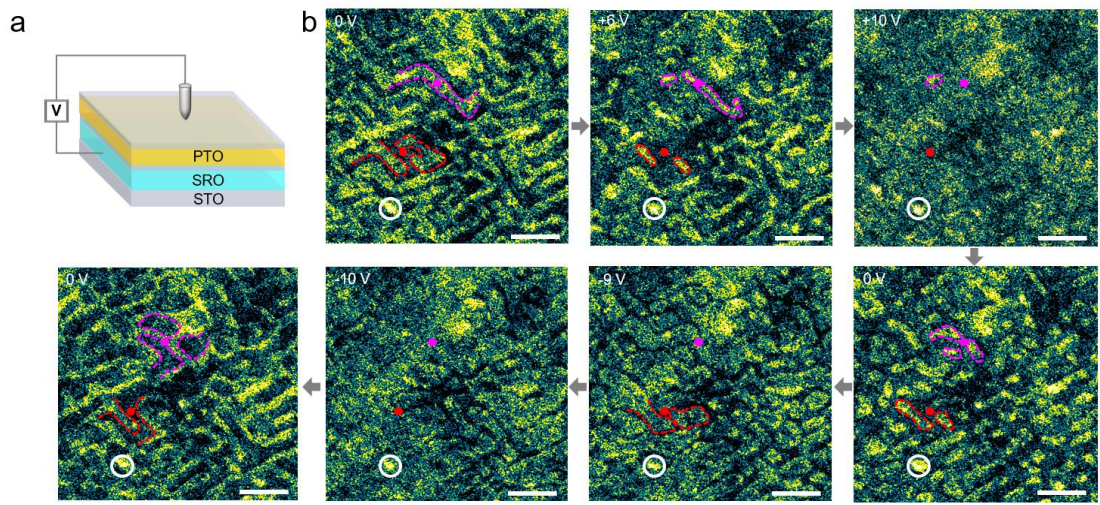

**Fig. 2. Electrical switching dynamics of skyrmion bubbles and stripe domains in a trilayer. a**, Schematic image showing the experimental set-up of in situ scanning transmission electron microscopy. Tungsten tip and SRO conductive layer are used as electrodes for applying an electric field to the film. **b**, Selected snapshots showing the evolution of stripe domains and skyrmion bubbles under a period of positive and negative bias loading. The red and purple dashed lines outline the characteristic domains, and the small dots of two colors are used to locate the domains. The white circle indicates the position of the marker for reference. Scale bars, 50 nm, each frame is processed using principal component analysis for better visualization.

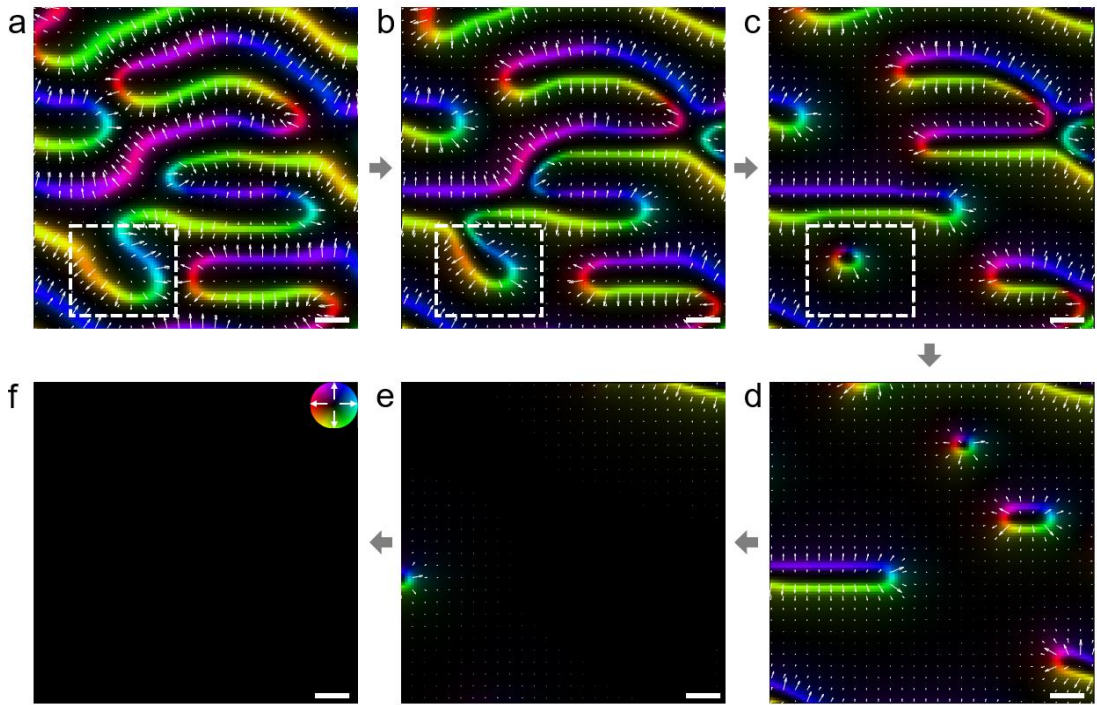

**Fig. 3. The evolution of coexisting structure in trilayer under positive electric field from phase-field simulation. a-f**, Electric field induced topological phase transition from skyrmion bubbles and stripe domains to trivial monodomain. Uniform saturated monodomain formed under a critical electric field of $E_{[001]}$ = 3400 kV/cm. The polar vectors and color mappings are depicted using the magnitude and direction of in-plane polarization, while the dark areas correspond to the out-of-plane polarization. The white dashed rectangles highlight the split of skyrmion bubbles. Scale bars, 5 nm.

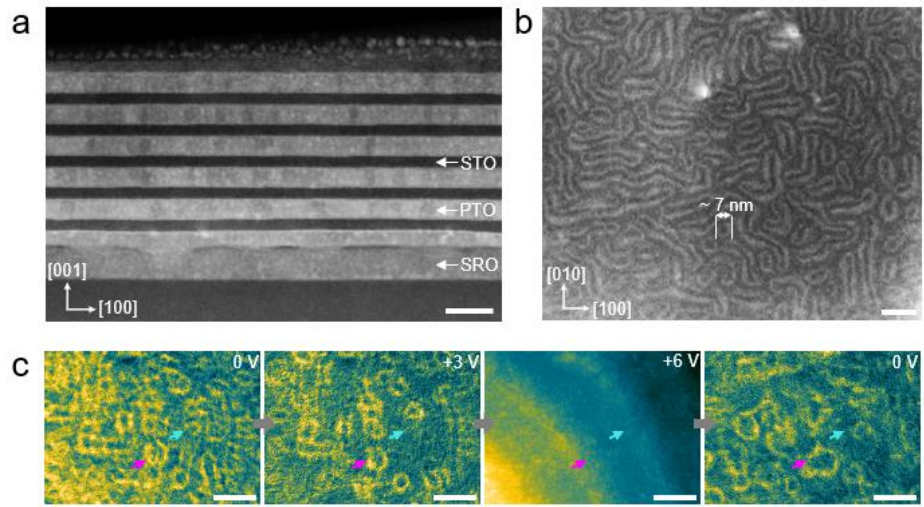

**Fig. 4. Electrical switching dynamics of skyrmion bubbles and stripe domains in a superlattice. a, b**, Cross-sectional (a) and planar view (b) STEM-MAADF images of a $[(SrTiO_3)_{11}/(PbTiO_3)_{22}/(SrTiO_3)_{11}]_6$ superlattice. Scale bars, 20 nm. **c**, Evolution of the skyrmion bubbles under a positive bias loading. Scale bars, 20 nm. The characteristic polar skyrmions are indicated by cyan and purple arrows respectively. For better visualization of polar skyrmions, the first (0 V), second (+3 V), and fourth (0 V) snapshots subtracted the third one (+6 V) which serve as the background.

# Supplementary Information for

# Dynamics of Polar Skyrmion Bubbles under Electric Fields

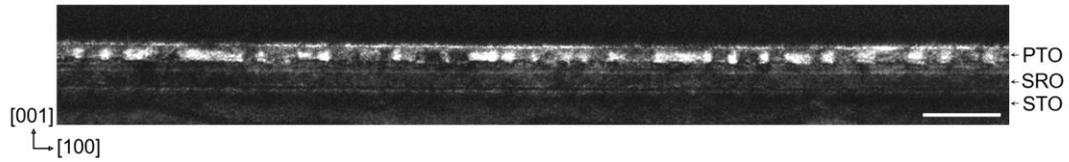

**Figure S1. Low-magnification dark field TEM image of a trilayer.** Cross sectional TEM dark-field image acquired by reflection with **g** = (200). Scale bar, 50 nm.

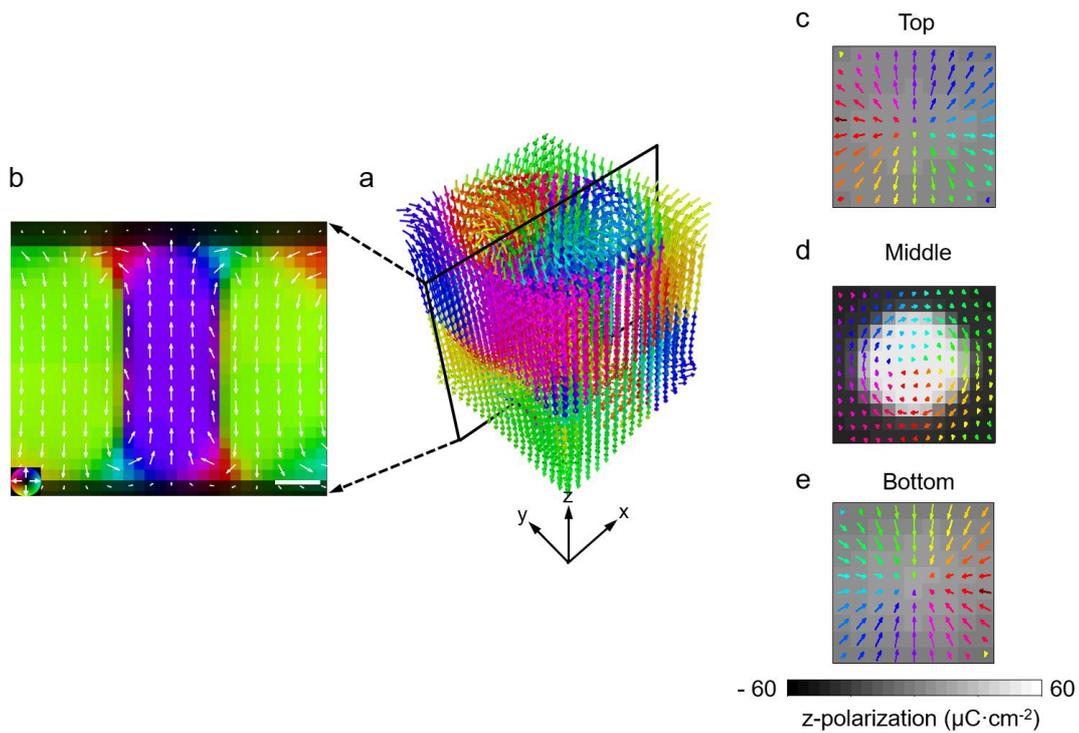

**Figure S2. Structure of a polar skyrmion bubble. a**, Three-dimensional texture of a skyrmion bubble. The color of the vectors corresponds to the polarization direction in the x-y plane. **b**, Cross-sectional polarization distribution of a skyrmion bubble. The white arrows denote the direction and magnitude of polarization, and the background color mapping represents the polarization direction. Scale bar, 2 nm. **c-e**, Polarization of the horizontal sections taken from the top (c), middle (d), and bottom (e) of a skyrmion bubble, with polar vectors showing the characteristics of central divergence, vortex, and central convergence, respectively. The color of polar vectors represents the polarization direction and the background intensity represents the magnitude of z-polarization.

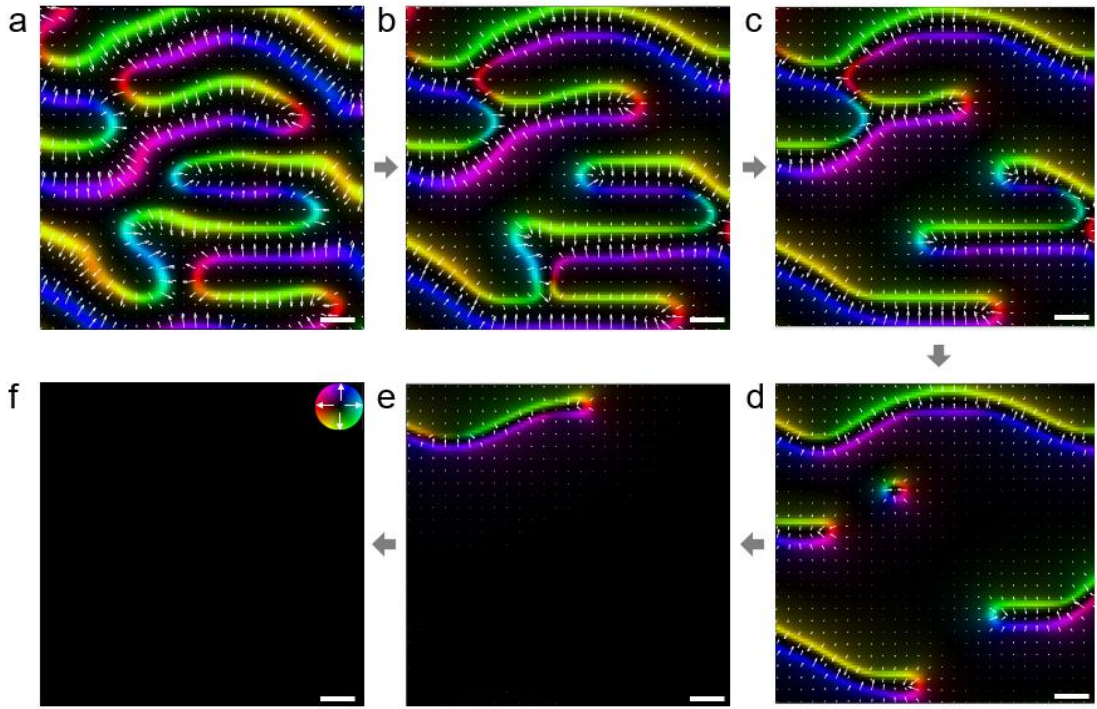

**Figure S3. The evolution of coexisting structure in trilayer under negative electric field from phase field simulation. a-f**, Electric field induced topological phase transition from skyrmion bubbles and stripe domains to trivial monodomain. Uniform saturated monodomain formed under a critical electric field of $E_{[001]}$ = -3400 kV/cm. The polar vectors and color mappings are depicted using the magnitude and direction of in-plane polarization, while the dark areas correspond to the out-of-plane polarization. Scale bars, 5 nm.

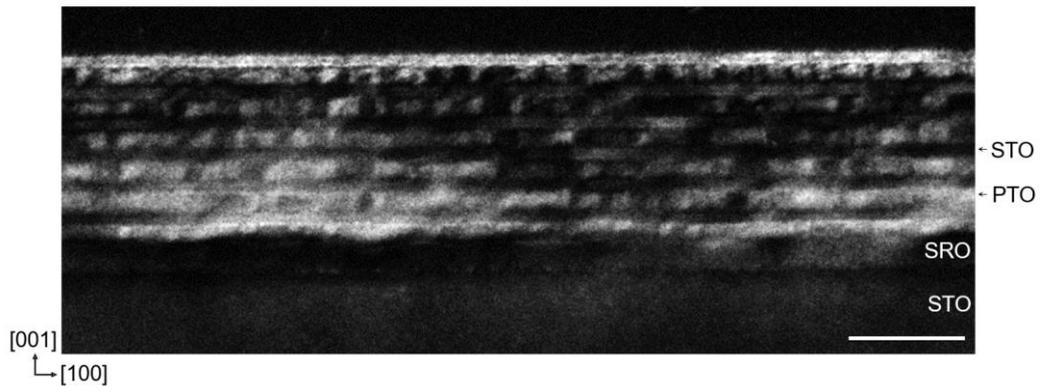

**Figure S4. Low-magnification dark field TEM image of a superlattice.** Cross sectional TEM dark-field image acquired by reflection with **g** = (200). Scale bar, 50 nm.

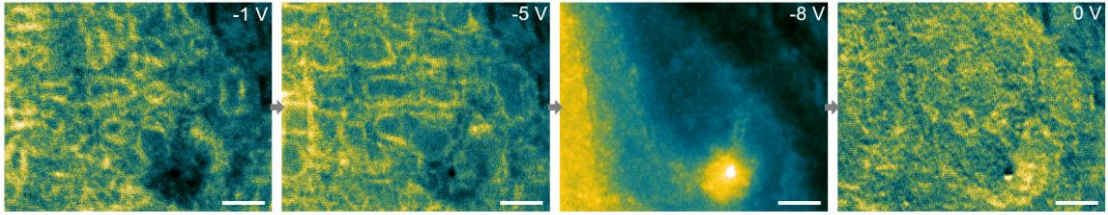

**Figure S5. Electrical switching dynamics of skyrmion bubbles in a superlattice.** Evolution of the skyrmion bubbles under a negative bias loading following the process shown in Fig. 4c. For better visualization of polar skyrmions, the first (-1 V), second (-5 V), and fourth (0 V) snapshots subtracted the third one (-8 V) which serve as the background. Scale bars, 20 nm.